\documentclass[aps,prx,twocolumn,longbibliography,groupedaddress]{revtex4-1}%
\usepackage{amsfonts}
\usepackage{amsmath,amssymb,epsfig,color}
\usepackage{float}
\usepackage{amsmath}
\usepackage{amssymb}
\usepackage{graphicx}%
\AtBeginDocument{%
    \newwrite\bibnotes
    \def\bibnotesext{Notes.bib}
    \immediate\openout\bibnotes=\jobname\bibnotesext
    \immediate\write\bibnotes{@CONTROL{REVTEX41Control}}
    \immediate\write\bibnotes{@CONTROL{%
    apsrev41Control,author="08",editor="1",pages="1",title="0",year="1"}}
     \if@filesw
     \immediate\write\@auxout{\string\citation{apsrev41Control}}%
    \fi
}%
\providecommand{\U}[1]{\protect\rule{.1in}{.1in}}
\providecommand{\U}[1]{\protect\rule{.1in}{.1in}}

\begin{document}
\title{Ising superconductivity and magnetism in NbSe$_{2}$}
\author{Darshana Wickramaratne}
\affiliation{NRC Research Associate, Resident at Center for Computational Materials
Science, U.S. Naval Research Laboratory, Washington, DC 20375, USA}
\author{Sergii Khmelevskyi}
\affiliation{
Research Center for Materials Science and Engineering, Vienna University of Technology, Karlsplatz 13, 1040 Vienna, Austria.
}
\author{Daniel F. Agterberg}
\affiliation{Department of Physics, University of Wisconsin, Milwaukee, WI 53201, USA}
\author{I.I. Mazin}
\affiliation{Department of Physics and Astronomy, George Mason University, Fairfax, VA
22030, USA}
\affiliation{Quantum Science and Engineering Center, George Mason University, Fairfax, VA 22030, USA}
\date{\today}

\begin{abstract}
Recent studies on superconductivity in NbSe$_{2}$ have demonstrated a large
anisotropy in the superconducting critical field when the material is reduced
to a single monolayer. Motivated by this recent discovery, we use density
functional theory (DFT) calculations to quantitatively address the
superconducting properties of bulk and monolayer NbSe$_{2}$. We demonstrate
that NbSe$_{2}$ is close to a ferromagnetic instability, and analyze our
results in the context of experimental measurements of the spin susceptibility
in NbSe$_{2}$. We show how this magnetic instability, which is pronounced in a
single monolayer, can enable sizeable singlet-triplet mixing of the
superconducting order parameter, contrary to contemporary considerations of
the pairing symmetry in monolayer NbSe$_{2}$, and discuss approaches as to how
this degree of mixing can be addressed quantitatively within our DFT
framework. Our calculations also enable a quantitative description of the
large anisotropy of the superconducting critical field, using DFT calculations
of monolayer NbSe$_{2}$ in the normal state.

\end{abstract}
\maketitle

\section{Introduction}

The transition metal dichalcogenides (TMD) exhibit an astonishing
variety of phenomena and phase transitions, which includes charge-density
waves (CDW) \cite{leroux2015strong,guster2019coexistence,xi2015strongly,ugeda2016characterization,bianco2020weak}, superconductivity
\cite{frindt1972superconductivity}, and magnetism \cite{bonilla2018strong}.
Bulk 2H-NbSe$_{2}$, which is one of the canonical transition metal
dichalcogenides, exhibits a rich phase diagram, which includes a
superconducting and a CDW phase \cite{leroux2015strong}. Superconductivity in
bulk NbSe$_{2}$ has been studied extensively both experimentally
\cite{moncton1977neutron,frindt1972superconductivity,foner1973upper,khestanova2018unusual,
noat2015quasiparticle}
and theoretically \cite{calandra2009effect, lian2018unveiling}, and the
superconducting transition temperature, $T_{c}$, has been experimentally
identified as $\sim$ 7 K \cite{foner1973upper}. While coupling between the
superconducting and CDW order parameters is certainly possible, it was found
to be weak in NbSe$_{2}$: the superconducting phase remains robust while the
CDW phase collapses as a function of increasing pressure
\cite{leroux2015strong} and as a function of increasing disorder
introduced by electron irradiation \cite{cho2018using}. Since NbSe$_{2}$ is also a layered van der Waals
material, this has inspired several studies of superconductivity in monolayer
NbSe$_{2}$
\cite{xi2016ising,sergio2018tuning,cho2020distinct,shaffer2020crystalline}
and
several intriguing proposals that seek to exploit proximity induced effects at
interfaces between monolayer NbSe$_{2}$ and magnetic materials
\cite{zhou2016ising,
glodzik2020engineering,hamill2020unexpected,kezilebieke2020topological}.
Furthermore, while NbSe$_2$ is thus far the canonical example of an Ising superconductor,
several theoretical studies on other monolayer TMDs such as TaS$_2$, TaSe$_2$ \cite{sergio2018tuning} and preliminary
experimental investigations on monolayer MoS$_2$ and MoTe$_2$ \cite{lu2015evidence,cui2019transport} have also shown indications of
hosting Ising superconductivity.

Monolayer NbSe$_{2}$, unlike the bulk structure, lacks inversion symmetry,
which leads to a large spin-orbit (SO) splitting of the states at the
\textrm{K}, and its inversion partner, {\rm K}$^{\prime}$, points
\cite{xiao2012coupled} (there is an additional Fermi surface with states
around $\Gamma$, which we will also discuss later in this study). The
magnitude of the SO-splitting is larger than the superconducting order
parameter. The zero-magnetic field $T_{c}$ of monolayer NbSe$_{2}$ is $\sim$ 3
K, \cite{sergio2018tuning,xi2016ising}, which is lower than the bulk $T_{c}$.
The combination of SO-coupling and broken inversion symmetry locks the
pseudospins near {\rm K} and {\rm K}$^{\prime}$ to be parallel to the $c$-axis of the
monolayer. Due to time-reversal symmetry, pseudospins at the {\rm K} and {\rm K}$^{\prime
}$ points are antiparallel, and their energies are degenerate. Hence, the
Cooper pairs that form completely break their rotational invariance in spin
space. This leads to a novel phenomenon, aptly named Ising superconductivity.
One key consequence of this unique pairing is that the superconducting phase
survives in the presence of in-plane magnetic fields that considerably exceeds
the Pauli limit \cite{sergio2018tuning,xi2016ising}.

Thus far, theoretical analyses of the superconducting pairing mechanism in
monolayer NbSe$_{2}$ \cite{mockli2020ising,he2018magnetic} have relied on
model descriptions of superconductivity in materials that lack inversion
symmetry \cite{youn2012role,frigeri:2004,smidman:2017}, loosely based on the
band structure calculated from first principles\cite{he2018magnetic}.  However,
a quantitative description of superconductivity in monolayer NbSe$_2$ is lacking. There
is also a lack of consistency between first-principles descriptions of
superconductivity in bulk NbSe$_{2}$ and experimental results.
State-of-the-art first-principles calculations that are usually very accurate
for superconductors where the pairing is entirely due to electron-phonon
coupling, overestimate $T_{c}$ in bulk NbSe$_{2}$ and isostructural NbS$_{2}$
\cite{Matt,heil2017origin} by a factor of $\sim3$ and the zero-temperature gap
by a factor of $\sim4$. Furthermore, experimental measurements of the spin
susceptibility, $\chi_{s}$, in bulk NbSe$_{2}$ \cite{iavarone2008effect}
find a $\chi_{s}$ $\sim$ 3$\times10^{-4}$ emu/mole, which, as we will show
later, considerably exceeds the bare bulk Pauli susceptibility, $\chi_{0}$.

Two plausible mechanisms that can be invoked to explain this discrepancy
between theory and experiment are the potential role of strong
electron-electron interactions and strong spin fluctuations. In
Ref.~\onlinecite{heil2017origin}, the authors suggested the overestimation of
$T_{c}$ in their first-principles calculations can be corrected by accounting
for a reduction in the effective mass induced by electron-electron
interactions, which they described within the $GW$ approximation. However,
reducing the effective mass by a factor of $(m^{\ast}/m)$ reduces the density
of states (DOS) by a factor of $(m^{\ast}/m)$ and increases the magnitude of
the electron-phonon matrix element by a factor of $(m^{\ast}/m)$. Indeed, the
DOS is proportional to the one-electron Green's function and the electron-phonon
matrix element is proportional to the derivative of the \textit{inverse} Green's
function. Since the electron-phonon coupling constant depends linearly on the
DOS and quadratically on the matrix element, reducing the effective mass
\textit{increases} the strength of the electron-phonon coupling. Hence, strong
electron-electron interaction effects alone do not provide a solution to this discrepancy.

The latter mechanism, which is the role of strong spin fluctuations in
NbSe$_{2}$, has thus far remained unaddressed. Fluctuations in the magnetic
moment and magnetic order have been shown to be a source of pairing, or
pair-breaking, of Cooper pairs in a number of other materials
\cite{boeri2010effects, fay1980coexistence}. Furthermore, strong spin
fluctuations can also lead to a sizeable Stoner renormalization of $\chi_{0}$.
To our knowledge, all theoretical studies of bulk and monolayer NbSe$_{2}$ at
their equilibrium lattice parameters have found the material to be
non-magnetic. However, calculations of monolayer NbSe$_{2}$ subject to tensile
strain exceeding 2\% have predicted a ferromagnetic ground state
\cite{xu2014tensile}.

This seeming lack of consistency between the various theoretical and
experimental results on bulk NbSe$_{2}$ reported in the literature indicates
there is still a need to explore the fundamental properties of NbSe$_{2}$. For
example, if spin fluctuations are operative in NbSe$_{2}$ it is unclear how
this may impact arguably the most interesting aspect of Ising
superconductivity in monolayer NbSe$_{2}$, which is the possibility of a
singlet-triplet mixed state. 
Singlet-triplet mixing of the superconducting order parameter 
has been attracting a lot of recent attention for a variety
of reasons which includes the ability to achieve spin supercurrents \cite{vorontsov2008surface,smidman:2017}
and the ability to drive superconducting topological transitions \cite{lu2008zero,smidman:2017}.
While Ising superconductivity is well understood
at the phenomenological level, it has never been described on a quantitative
level using first-principles calculations. This also precludes a quantitative
description of Ising superconductivity, which has been observed in monolayers
of several other transition metal dichalcogenides beyond NbSe$_{2}$.
\cite{sergio2018tuning,lu2015evidence,cui2019transport}.

In this paper we will demonstrate that, indeed, bulk and especially monolayer
NbSe$_{2}$ are close to a magnetic instability. We will \textquotedblleft
translate\textquotedblright\ the existing model theory that has been developed
to analyze superconductivity in materials that lack inversion symmetry and
bands split by SO-interaction \cite{frigeri2004spin}, such as monolayer
NbSe$_{2}$, into DFT parlance, which allows us to develop a quantitative
theory of the critical field anisotropy in this material. We use
collinear and non-collinear fixed-spin moment calculations to determine
the spin susceptibility of bulk and monolayer NbSe$_{2}$. Finally, we will use
the insights obtained from our calculations to discuss possible ramifications
on the superconducting order parameter in NbSe$_{2}$, in particular, the
factors that control the scale of the triplet admixture to the order parameter.
This knowledge is the first step towards a comprehensive quantitative theory
of Ising superconductivity in monolayer NbSe$_2$.

\section{Electronic structure}

\begin{figure}[ptb]
\includegraphics[width=8.5cm]{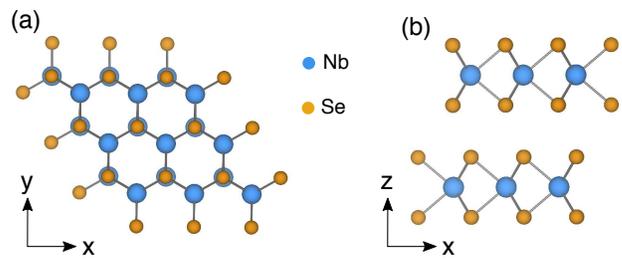}\caption{(a) Crystal structure of bulk
NbSe$_{2}$ illustrating the (a) top view and (b) side view of the structure.
The $x,y$ and $z$ axes denote the cartesian axes.}%
\label{fig:struct}%
\end{figure}

The bulk unit cell of 2H-NbSe$_{2}$ consists of two monolayers of NbSe$_{2}$
(Fig.\ref{fig:struct}(a)) where in a single monolayer the Nb atoms are in a
trigonal prismatic coordination with the Se atoms. The Nb atoms in one of the
monolayers are vertically above the Se atoms of the second monolayer (cf.
Fig.~\ref{fig:struct}(b)), which leads to a center of inversion that is
between the two monolayers of the unit cell. The calculated bulk lattice
constants, $a$=3.449 \AA ~, and $c$=12.550 \AA , are in agreement with
experimental reports of the lattice constants of bulk NbSe$_{2}$
\cite{marezio1972crystal}. 

The trigonal crystal field splits the $4d$-states of Nb$^{4+}$ into three
different groups: $d_{z^{2}}$, [$d_{x^{2}-y^{2}}$, $d_{xy}$] and [$d_{xz}%
$,$d_{yz}$]. The bulk band structure of NbSe$_{2}$, which has been studied
extensively \cite{mattheiss1973energy,johannes2006fermi}, has three bands that
cross the Fermi level. Two of the bands are derived from Nb $d$-states and the
third band is derived from Se $p_{z}$ states. Spin-orbit interaction leads to
a mixing of the Se $p$ and Nb $d$-states along the $\Gamma$-\textrm{K}%
-\textrm{M} path of the Brillouin zone, but does not lead to SO-splitting. The
density of states at the Fermi level, $N(E_{F})$ is 2.7 states$/(eV\cdot$cell),
which leads to a bulk Pauli susceptibility, $\chi_{0}$ = 0.87$\times$10$^{-4}$
emu/mole, which is a factor of $\sim$ 3.5 lower than the experimentally
reported spin susceptibility of bulk NbSe$_{2}$ \cite{iavarone2008effect},
which suggests that spin fluctuations are operative in NbSe$_{2}$.

\begin{figure}[ptb]
\includegraphics[width=8.5cm]{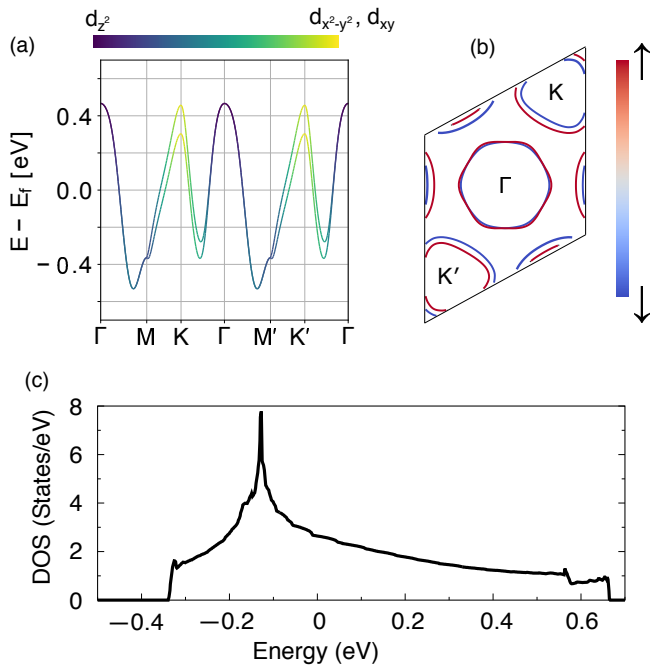}\caption{(a) Band structure of
monolayer of NbSe$_{2}$. The color along each band denotes the relative Nb
$d_{z^{2}}$, $d_{x^{2}-y^{2}}$ and $d_{xy}$ character along the high-symmetry
path, according the color bar above the plot. (b) Cross section of the Fermi
surface of monolayer NbSe$_{2}$. Red denotes bands that have pure $m_{z}$=1
character while blue denotes bands that have pure $m_{z}$=$-1$ character. (c)
Density of states of monolayer NbSe$_{2}$. All of the calculations include
SO-coupling.}%
\label{fig:ek}%
\end{figure}

For the case of the monolayer, there is one band that crosses the Fermi level
several times, leading to three Fermi contours, one contour around the
$\Gamma$-point and two contours around {\rm K} and {\rm K}$^{\prime}$ (these are related
by inversion symmetry). At $\Gamma$, the band character is Nb $d_{z^{2}},$
with a minor admixture of Se $p_{z}.$ As the band progresses toward {\rm K} or
{\rm K}$^{\prime},$ this leads to a larger admixture of Nb $d_{x^{2}-y^{2}}$ and
$d_{xy}$ orbitals in addition to a minor contribution of Se $p_{xy}$ states.
The states at the {\rm K} and {\rm K}$^{\prime}$ contours are comprised entirely of Nb
[$d_{x^{2}-y^{2}}$, $d_{xy}$] states. In the absence of SO-interaction, this
band is spin-degenerate. When we allow for SO-interaction, the lack of a
center of inversion in the monolayer leads to SO-splitting everywhere except
along the $\Gamma$-M line (cf. Fig~.\ref{fig:ek}(a)), consistent with 
prior calculations of the band structure of monolayer NbSe$_2$ \cite{silva2016electronic}. 

To understand why the pseudospin state does not have an in-plane component and
why the splitting is small near $\Gamma$, it is instructive to rationalize
this splitting from the band structure point of view. If we neglect the minor
admixture of Se $p$-states to the bands that cross the Fermi level in
monolayer NbSe$_{2}$, a state at a given wave vector \textbf{k} can be defined
as follows:
\begin{equation}
\left\vert {\phi}\right\rangle =\eta\left\vert d_{x^{2}-y^{2}}\right\rangle
+\beta\left\vert d_{xy}\right\rangle +\gamma\left\vert d_{z^{2}}\right\rangle
, \label{abg}%
\end{equation}
where $|\eta|^{2}+|\beta|^{2}+|\gamma|^{2}=1$. Note that $d_{z^{2}}$ corresponds to
$\left\vert l,m\right\rangle =\left\vert 2,0\right\rangle ,$ $d_{x^{2}-y^{2}}$
to $(\left\vert 2,2\right\rangle +\left\vert 2,-2\right\rangle )\sqrt{2}$, and
$d_{xy}$ to $(\left\vert 2,2\right\rangle -\left\vert 2,-2\right\rangle
)/\mathrm{i}\sqrt{2}$.
Accounting for spin, the Hamiltonian at each $\mathbf{k}$ point is a
($2\times2$) matrix, and, by virtue of the $z/-z$ mirror symmetry, does not
include contributions from the $\left\vert 2,\pm1\right\rangle $ orbitals.
Thus, the nondiagonal matrix elements $L_{\pm}$ are zero. However, it is easy
to show that the diagonal element $L_{z}=2(\eta\operatorname{Im}\beta
-\beta\operatorname{Im}\eta).$ One phase can always be selected as real, for
instance, $\eta,$ then $L_{z}=2\eta\operatorname{Im}\beta.$

In the centrosymmetric bulk 2H-NbSe$_{2}$, $\beta$ can also be chosen to be
real, and there is no SO-induced spin-splitting (but there is splitting due to
doubling of the unit cell). In the monolayer, $\beta$ is complex everywhere
except the $\Gamma-$M direction, and therefore the diagonal elements of this
($2\times2$) matrix have opposite signs, $\pm\lambda\eta\operatorname{Im}%
\beta,$ where $\lambda$ measures the strength of the SO-coupling.
Consequently, the splitting is small around the $\Gamma$ pocket (the maximum
splitting at this Fermi-surface is $\sim$70 meV, which occurs where it
intersects with the $\Gamma-${\rm K} and $\Gamma-${\rm K}$^{\prime}$ lines), where
$|\gamma|^{2}\gg|\eta\beta|,$ but sizeable ($\sim$ 150 meV) on the \textrm{K}
and \textrm{K$^{\prime}$ contours}, where $|\gamma|^{2}\ll|\eta\beta|.$ Due to
the absence of nondiagonal coupling, the pseudospin-split states are also pure
$S_{z}=\pm\frac{1}{2}$ spin states and the direction of the pseudospin flips
between the {\rm K} and {\rm K}$^{\prime}$ valleys as illustrated in Fig.~\ref{fig:ek}(b).
As we will discuss later, pseudospin states are no longer pure $S_{z}$ states
when an external in-plane magnetic field is applied.

\section{Magnetism without spin-orbit coupling}

To quantitatively address the role of spin fluctuations we calculate the spin
susceptibility, $\chi$, by considering the effect that a uniform external
magnetic field, $H$, has on the magnetic moment, $m$, in NbSe$_{2},$ via the
Zeeman interaction, where $\chi=\partial m/\partial H$. In practice, one way
to simulate the effect of a magnetic field within a first-principles calculation, is to apply a
constraint on the magnetization and compute the total energy, $E$, as a
function of the magnetic moment, $m$. This \textquotedblleft fixed-spin
moment\textquotedblright\ (FSM) approach allows us to define $\chi$ as
$\chi=(\partial^{2}E/\partial m^{2})^{-1}$. Similarly, non-collinear FSM
calculations that include SO-interaction let us determine the change in total
energy and in turn $\chi$ for directions parallel to the $c$-axis
($\langle001\rangle$) and perpendicular to the $c$-axis ($\langle100\rangle$)
of NbSe$_{2}$.

\begin{figure}[ptb]
\textrm{\includegraphics[width=8.5cm]{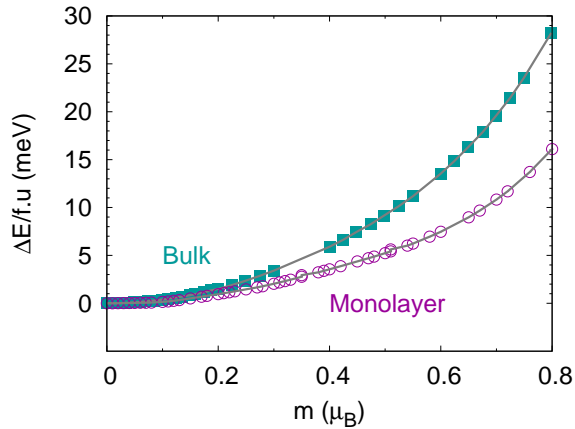}}\caption{Change in
total energy per NbSe$_2$ formula unit (f.u.) of the 
bulk ($\square$) and monolayer ($\circ$) structures with
respect to the non-magnetic state as a function of magnetic moment obtained
from collinear fixed-spin moment calculations. The grey solid lines are
a fit to Eq.~\ref{eq:chi}.}%
\label{fig:fsm}%
\end{figure}

The collinear and non-collinear (along $\langle
001\rangle$ and $\langle100\rangle$) FSM calculations 
result in the same qualitative trends;
the total energy increases monotonically with respect to the total energy of
the non-magnetic state as a function of increasing magnetic moment. As an
example, we illustrate the results of our collinear FSM calculations
for bulk and monolayer NbSe$_{2}$ in Fig.~\ref{fig:fsm}. If we express the expansion of the DFT total energy as
\begin{equation}
E(m)=a_{0}+a_{1}m^{2}+a_{2}m^{4}+a_{3}m^{6}+a_{4}m^{8}+\dots\label{eq:chi}%
\end{equation}
we can determine the static spin-susceptibility, $\chi$, for low values of $m$
by using the coefficient $a_{1}$ obtained by fitting the data in
Fig.~\ref{fig:fsm} to Eq.~\ref{eq:chi}. The results are summarized in Table
\ref{tab:chi}.

\begin{table}[ptb]
\textrm{
\begin{tabular}
[c]{ccc}\hline\hline
Spin susceptibility & \multicolumn{1}{l}{Bulk} & \multicolumn{1}{l}{Monolayer}%
\\[0.5ex]
& \multicolumn{2}{l}{[10$^{-4}$ emu/mole]}\\[0.5ex]\hline
Collinear & 4.28 & 6.81\\[0.5ex]%
{$\langle001\rangle$} & 4.20 & 7.29\\[0.5ex]%
{$\langle100\rangle$} & 4.23 & 7.40\\[0.5ex]%
Experiment & $\sim$3 & $-$\\\hline\hline
\end{tabular}
}\caption{Spin susceptibility of the bulk and monolayer structures obtained
from collinear and non-collinear calculations with the spin quantization
axis parallel to the $c$-axis, $\langle001\rangle$, and the spin quantization
axis along the $x$ direction, $\langle100\rangle$. The calculated
susceptibility along $\langle100\rangle$, $\langle010\rangle$, and
$\langle110\rangle$ are equivalent. The experimental value of $\chi$,
obtained from Ref.~\onlinecite{iavarone2008effect}, is for a
magnetic field applied parallel to the $c$-axis ($\langle001\rangle$).
The experimental value of $\chi$ for monolayer NbSe$_2$ to our knowledge has not been reported yet.}
\label{tab:chi}%
\end{table}

Based on our FSM calculations, we can draw the following conclusions. First,
DFT repoduces the experimentally observed bulk susceptibility reasonably well,
only slightly overestimating it compared to the experimental value. This
overestimation in the calculated value of $\chi$ is known to occur in
itinerant metals close to a magnetic instability, and is due to a
fluctuational reduction of the mean-field DFT
moment\cite{moriya2012spin,larson2004magnetism}. Applying Moriya's theory
\cite{moriya2012spin} to NbSe$_{2}$, we can estimate the average magnitude of
spin fluctuations as $\xi \sim0.28$ $\mu_{B}$. To put this into context, the
average magnitude of spin fluctuations in palladium (Pd), a known
superparamagnet (which at some point was considered a candidate for triplet
superconductivity \cite{fay1980coexistence}) was calculated (in LDA, as
opposed to our GGA calculation) to be $\xi \sim0.15$ $\mu_{B}$%
\cite{larson2004magnetism}.

This is also consistent with our disordered local moment calculations (DLM),
where the energy cost of creating a local spin fluctuation with an amplitude
of $\sim0.2$ $\mu_{B}$ is nearly twice higher in Pd than in NbSe$_{2}.$ This
does not mean that NbSe$_{2}$ is closer to ferromagnetism compared to Pd. The
molar spin susceptibility of NbSe$_{2}$ is a factor of two lower compared to
Pd. However, it does mean spin fluctuations in NbSe$_{2}$ are soft over a
large part of the Brillouin zone. Furthermore, it is important to note the
susceptibility of the monolayer structure is $\sim$50\% larger than that of
the bulk, which indicates that spin fluctuations are stronger in a monolayer.
Indeed, this is consistent with monolayer NbSe$_{2}$ having a lower
superconducting transition temperature compared to bulk NbSe$_{2}$
\cite{xi2016ising}.

\begin{figure}[ptb]
\includegraphics[width=8.5cm]{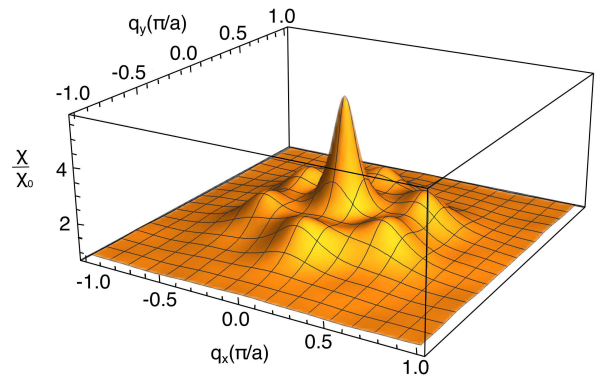}\caption{\textbf{q}-dependent Stoner
renormalization factor of bulk NbSe$_{2},$ obtained from disordered local
moment calculations.}%
\label{fig:Jq}%
\end{figure}

To verify that NbSe$_{2}$ is indeed close to a \textit{ferro}magnetic
instability, we also calculated the exchange coupling between fluctuating
moments within the DLM formalism (the calculation details are similar to
methods used in Ref.~\cite{kim2017anisotropy}) for bulk NbSe$_{2}$. 
The exchange coupling is largely dominated by the nearest-neighbor coupling,
which we find to be ferromagnetic. In order to transform the exchange
interactions into reciprocal space, $J_{0}(\mathbf{q)}$, we have defined
$\chi_{RPA}(\mathbf{q)}=c_{0}/[1-c_{0}J_{0}(\mathbf{q})],$ where $c_{0}$ is a
constant of the order of $N(0).$ In Figure \ref{fig:Jq} we plot the
renormalization factor $1/[1-c_{0}J(\mathbf{q})]$, using $c_{0}=3.57$ eV$^{-1}%
$, which was chosen so as to have the renormalization at $\mathbf{q} = 0$ be
approximately consistent with a renormalization factor of 4.9.
The peak near $\mathbf{q} = 0$ in Figure \ref{fig:Jq} indicates that
the system is close to a ferromagnetic instability.

The DLM calculations allow us to determine the $\mathbf{q}$-dependence
of $\chi$ and offers {\it qualitative} insight into the magnitude of $\chi$.
For accurate {\it quantitative} values of $\chi$ we refer
the reader to Table~\ref{tab:chi}, which are obtained using FSM calculations.

\section{Magnetism with spin-orbit coupling}

Having demonstrated that bulk and monolayer NbSe$_{2}$ are indeed close to a
ferromagnetic instability, we now focus on the effect of SO-coupling, and
specifically on the response to an external magnetic field applied parallel to
the $c$-axis and perpendicular to the $c$-axis. First, it is evident the
values of $\chi$ reported in Table \ref{tab:chi} from our non-collinear
calculations are isotropic along $\langle001\rangle$ and $\langle100\rangle$,
within a few percent, for both the bulk and monolayer structures, as opposed
to the susceptibility in the superconducting state.

To understand this, let us analyze how the bands that cross the Fermi level
evolve as a function of the magnitude and direction of an applied magnetic
field. In the absence of SO-interaction, the states at $\Gamma$, {\rm K} and
{\rm K}$^{\prime}$ are degenerate. The Zeeman interaction, regardless of the
direction of the field, splits the bands by approximately the same magnitude,
$\pm H,$ where $H$ is the Stoner-enhanced external field (we have absorbed the
Bohr magneton in the units of $H)$. Indeed, in our calculations, the splitting
at $\Gamma$, {\rm K} and {\rm K}$^{\prime}$ increases linearly and by approximately the
same amount as a function of increasing magnetic moment. The magnitude of the
Fermi surface splitting in reciprocal space is $\delta k_{F}(\mathbf{k)=}%
2H/v_{F}(\mathbf{k)}$.
The area between the spin-split contours will determine the total
magnetization for a given magnitude of the magnetic field, and will be
$2N_{\uparrow}H=N_{\uparrow\downarrow}H,$ where $N_{\uparrow}$ is the total
number of states with pseudospins along $\hat{z}$ and $N_{\uparrow\downarrow}$
is the total number of states with pseudospins along $\hat{z}$ and $-\hat{z}$.

We now define the following generic Hamiltonian with SO-interaction for a
given point on the Fermi surface of monolayer NbSe$_{2}$ subject to an
external magnetic field $H_{z}$ along $\hat{z}$ (parallel to the $c$-axis),
where the spin-quantization axis is along $\hat{z}.$
\begin{equation}
\mathfrak{H(}\mathbf{k)}=\left(
\begin{array}
[c]{cc}%
\varepsilon_{\mathbf{k}}\pm\lambda_{\mathbf{k}}+H_{z} & 0\\
0 & \varepsilon_{\mathbf{k}}\mp\lambda_{\mathbf{k}}-H_{z}%
\end{array}
\right)  \label{eq:hz}%
\end{equation}
Based on Eq. \ref{abg}, $\lambda_{\mathbf{k}}=\lambda\eta_{\mathbf{k}%
}\operatorname{Im}\beta_{\mathbf{k}},$ and the matrix is in the $S_{z}$ spin
space. Inversion in the momentum space changes the upper sign to the lower
sign for $\pm\lambda$ and $\mp\lambda$. The splitting between the Fermi
contours for a given pseudospin along $+\hat{z}$ will increase by
approximately the same magnitude as the Fermi contours of the majority spin
channel in the case of the collinear calculations. The splitting of the
Fermi contours for pseudospins along $-\hat{z}$ will decrease by the same
magnitude. For instance, if the Fermi contour splitting around {\rm K} increases by
$2H_{z},$ the splitting around {\rm K}$^{\prime}$ will decrease by the same amount.
Hence, the total magnetization that is induced in terms of the population of
pseudospins, to lowest order in $\lambda$, will be exactly the same as
determined by our collinear calculations.

\begin{figure}[ptb]
\textrm{\includegraphics[width=8.5cm]{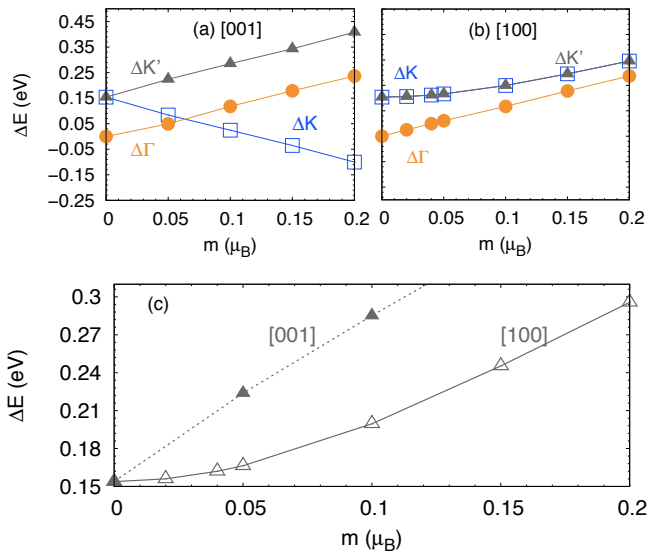}}\caption{Splitting at
$\Gamma$ ($\circ$), {\rm K} ($\square$) and {\rm K}$^{\prime}$ ($\bigtriangleup$) obtained
with non-collinear calculations as a function of the magnetic moment on Nb for
magnetic moments (a) parallel to the $c$-axis, $\langle001\rangle$ and (b)
perpendicular to the $c$ axis, $\langle100\rangle$ for monolayer NbSe$_{2}$.
The magnitude of the splitting for magnetic moments along $\langle010\rangle$
are similar to the results along $\langle100\rangle$ illustrated in (b).
The solid line in (a) and (b) for each plot serves as a guide to the eye.  
(c)  Splitting at {\rm K}$^{\prime}$ for magnetic moments on Nb parallel to the $c$-axis 
([001] dotted line) illustrating the linear dependence on $m$ 
and magnetic moments perpendicular to the $c$-axis ([100] solid line)
illustrating the quadratic dependence on $m$.
The results are the same at {\rm K} if one considers the magnitude of the change
in $\Delta E$ versus $m$.}
\label{fig:splitting}%
\end{figure}

In the non-collinear DFT calculations, a pseudospin state is formally a
combination of both spins, so we introduce an effective $g$ factor, which
describes the difference between the pure spin susceptibility and the
\textquotedblleft pseudospin\textquotedblright\ susceptibility. For a magnetic
field along $\hat{z}$, the $g$ factor is exactly 2, so we expect $\chi$ along
$\langle001\rangle$ to be very similar to $\chi$ obtained from collinear
calculations -- which is consistent with our calculations in Table
\ref{tab:chi}. The splitting of the states at $\Gamma$, {\rm K} and {\rm K}$^{\prime}$ for
magnetic moments along $\hat{z}$ is illustrated in Fig.~\ref{fig:splitting}(a)
and they indeed change linearly with respect to the magnetic moment.

For an in-plane magnetic field along $\hat{x}$ (perpendicular to the
$c$-axis), $H_{x}$, the Zeeman interaction, $S_{x}H_{x}=(S_{+}+S_{-})H_{x}/2$,
couples to the off-diagonal components of the spin-orbit Hamiltonian as
follows:
\[
\mathfrak{H(}\mathbf{k)}=\left(
\begin{array}
[c]{cc}%
\varepsilon_{\mathbf{k}}\pm\lambda_{\mathbf{k}} & \pm H_{x}/2\\
\mp H_{x}/2 & \varepsilon_{\mathbf{k}}\mp\lambda_{\mathbf{k}}%
\end{array}
\right)  .
\]
To linear order in $H_{x}$, the splitting between the Fermi contours at {\rm K} and
{\rm K}$^{\prime}$ will not change. However, the wave functions change, and thus the
effective $g$-factor will deviate linearly from 2. For example, applying 
standard perturbation theory to the pseudospin $\left\vert
+\right\rangle $ states, gives:
\begin{align}
\left\vert +\right\rangle  &  =\left\vert \uparrow\right\rangle \pm\frac
{H_{x}/2}{2\lambda_{\mathbf{k}}}\left\vert \downarrow\right\rangle \nonumber\\
\left\langle +\left\vert\frac{\sigma_{+}+\sigma_{-}}{2}\right\vert+\right\rangle  &  =\pm
\frac{H_{x}/2}{\lambda_{\mathbf{k}}} \label{g}%
\end{align}

Pseudospin $\left\vert -\right\rangle $ states will acquire the opposite
magnetization, and their $g$ factor will be reduced by the same amount. We now
observe that the total pseudo-moment around the {\rm K} point will be proportional
to the area between the split concentric Fermi contours, $\pm N_{K}\lambda
_{K}$, where $N_{K},$ is the density of states for this contour at {\rm K} and
$\lambda_{K}$ is the average splitting at this contour. Around the {\rm K}$^{\prime
}$ contour, the area between the split concentric Fermi contours is $\mp
N_{K}\lambda_{K}$, and this total \textit{pseudo}-moment does not depend on
$H_{x}.$ Multiplying it by the difference in the $g$ factors of Eq. \ref{g},
which deviate from 2 by the same amount, but in the opposite directions, we
get a spin susceptibilty of $\chi_{\langle100\rangle}\approx N_{tot}%
=\chi_{\mathrm{Pauli}}$, where $N_{tot}$ is the total density of states. Thus,
no anisotropy in the spin susceptibility appears in the lowest order of the
SO-coupling. DFT calculations fully conform with this description: the
splitting of the one-electron energies at {\rm K} and {\rm K}$^{\prime}$ is quadratic with
respect to magnetic moments oriented along $\hat{x}$ (cf.
Fig.~\ref{fig:splitting}(b)).

Within our considerations of the Zeeman interaction, $H$ is the total magnetic
field, which includes the Stoner renormalization. Within DFT, the RPA
is exact, since one can write the total DFT exchange-correlation
energy, $E_{xc}$, in an external magnetic field
as\cite{krasko1987metamagnetic,andersen1977magnetic}:
\begin{equation}
E_{xc}=\frac{m^{2}}{4}\left(\frac{1}{N_{\uparrow}}-I\right),
\end{equation}
where $I=\delta^{2}E_{xc}/dm^{2}$ is the DFT Stoner factor, which in the DFT
language combines the diagonal (Hubbard $U)$ and off-diagonal (Hund's $J)$
interactions. Indeed,
\begin{align}%
\begin{split}
\chi_{DFT}  &  =\chi_{0}/(1-\chi_{0}I)\\
H  &  =H_{ext}/(1-\chi_{0}I)
\end{split}
\end{align}
where $\chi_{0}$ is the bare Pauli susceptibility.

\section{Magnetism and superconductivity}

\label{sec:results-mag} Within this framework, it is especially easy to
address the effect of superconductivity on the spin susceptibility. Indeed,
the opening of the superconducting gap, $\Delta$, only affects states that are
close to the Fermi surface. Since the spin susceptibility parallel to the
$c$-axis, $\chi_{\langle001\rangle}$, is determined entirely by the shift of
the Fermi contours as a function of an increasing magnetic field
(Fig.~\ref{fig:splitting}(a)), the spin susceptibility is suppressed by
superconductivity in exactly the same way as without SO-coupling. In contrast,
the spin susceptibility perpendicular to the $c$-axis, $\chi_{\langle
100\rangle}$, as we just saw, is defined by the states removed from the Fermi
level by $\sim\lambda_{\mathbf{k}}\gg\Delta,$ and as a result is not affected
by $\Delta$. Thus, the thermodynamic critical field, $H_{C0},$ which is
determined by the free energies in the normal and superconducting state as
\begin{equation}
F_{n}-F_{s}\sim\Delta^{2}N_{\uparrow\downarrow}(0)/2=(\chi_{n}-\chi_{s}%
)H_{C0}^{2}/2
\end{equation}
behaves conventionally for magnetic fields parallel to the $c$-axis, but is
essentially infinite for magnetic fields perpendicular to the $c$-axis.

However, if one examines the SO-splitting for the Fermi contour around the
$\Gamma$-point, we find that it is nodal along the $\Gamma$-M and the $\Gamma
$-M$^{\prime}$ line, which makes $H_{C0}$ finite, but, greatly enhanced for
magnetic fields perpendicular to the $c$-axis. Figure~\ref{fig:ek}(b),
illustrates the calculated splitting due to SO-coupling along the $\Gamma$
contour is finite and has nodes along all $\Gamma-$ M and $\Gamma-$M$^{\prime
}$ directions. The magnitude of this splitting is low, but, along the
antinodal line $\Gamma-$K, the maximum splitting is still larger than the
superconducting gap. In the Appendix (Sec.~\ref{sec:app-gamma}), we will
derive an analytical expression, which generalizes the considerations
presented above onto a SO-coupled nodal case of the $\Gamma$-centered Fermi contour.

\section{Singlet and triplet superconductivity: DFT point of view}
\label{sec:results-mixing} We now use the above considerations to determine
the symmetry of possible pairing interactions for monolayer NbSe$_2$. 
The fact that NbSe$_{2}$ lacks
inversion symmetry formally allows for parity mixing, but it has been argued
\cite{shaffer2020crystalline} that the triplet component must be vanishingly
small. As we will discuss below, this is not necessarily true. Strong
spin-fluctuations, which we have demonstrated to be operative in NbSe$_{2}$,
and/or a particular structure of the electron-phonon coupling have the ability
to generate a sizeable triplet component.

\begin{figure}[ptb]
\textrm{\includegraphics[width=8.5cm]{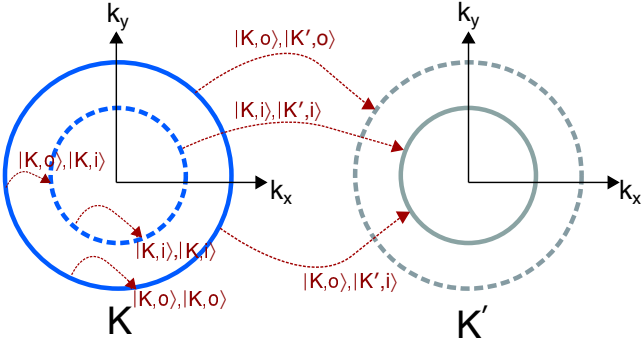}}\caption{Schematic
illustration of the inner and outer contours of the Fermi surface around the K
and K${\prime}$ points that cross the Fermi level (cf. Fig ~\ref{fig:ek}(b)).
Solid circles represent pseudospin $\left\vert \uparrow\right\rangle $ states
and dotted circles represent pseudospin $\left\vert \downarrow\right\rangle $
states. The possible pairing interactions due to phonons or spin fluctuations
between the pseudospin states are denoted with red dotted arrows. The relative
signs of these interactions are summarized in Table ~\ref{tab:pairs}.
Subscript $o$ refers to the outer contour and the subscript $i$ refers to the
inner contour at a given valley. }%
\label{fig:k_schem}%
\end{figure}

In the spirit of band theory, we consider the one-electron Hamiltonian to be
fully diagonalized \textit{before} we consider superconducting pairing. First,
we only consider the {\rm K} and {\rm K}$^{\prime}$ contours. The Cooper pairs at these
contours are comprised of states that reside on either the inner or outer
contours at {\rm K} and {\rm K}$^{\prime}$. We assume that the outer contour around {\rm K} has
spin-up, and the inner contour has spin-down states, which we denote as
[$\left\vert K,o,\uparrow\right\rangle $,$\left\vert K,i,\downarrow
\right\rangle $] The contours at {\rm K}$^{\prime}$ are degenerate in energy with
the contours at {\rm K} (the states at {\rm K} and {\rm K}$^{\prime}$ are related by
time-reversal symmetry) and is given by [$\left\vert K^{\prime},o,\downarrow
\right\rangle $,$\left\vert K^{\prime},i,\uparrow\right\rangle $].
No other combinations are allowed.  
Schematically, these four contours can be
represented as two pairs of concentric rings as depicted in
Fig.~\ref{fig:k_schem}.
 In this
basis, the anomalous averages that appear in the problem are
\begin{equation}%
\begin{split}
d_{o,\mathbf{k}}  &  =\overline{\left\vert K,o,\uparrow\right\rangle
\left\vert K^{\prime},o,\downarrow\right\rangle }\\
d_{i,\mathbf{k}}  &  =-\overline{\left\vert K,i,\downarrow\right\rangle
\left\vert K^{\prime},i,\uparrow\right\rangle }%
\end{split}
\label{eq:d0,di}%
\end{equation}

Notice that we introduced a minus sign in
Eq.~\ref{eq:d0,di} for $d_{i,\mathbf{k}}$, this is to ensure that a usual
spin-singlet state is given by $d_{i,\mathbf{k}}=d_{o,\mathbf{k}}$ and that
the interactions we discuss later reduce to the usual interactions when no
spin-orbit coupling is included. 
Indeed, one is free to define the relative phase between the
superconducting order parameters at different momenta, without changing the
superconducting state or any observables. Usually there is one logical
choice and it is universally used. 
For the case of NbSe$_2$ it is important to note that the relative sign between
the order parameter at K and K$^{\prime}$ is not uniquely defined.
Hence, one needs to be careful when defining the phase convention
for a given pairing interaction, as it can have an instrumental impact 
as has been shown for other materials \cite{parker2009coexistence}.

Since a singlet pair is defined as $(\left\vert \uparrow\downarrow
\right\rangle $ $-\left\vert \downarrow\uparrow\right\rangle )/\sqrt{2}$ and
the triplet pair is defined as $(\left\vert \uparrow\downarrow\right\rangle $
$+\left\vert \downarrow\uparrow\right\rangle )/\sqrt{2}$, the order parameter
on the outer contour, $\Delta_{o}$, derived from the anomalous average
$d_{o},$ is $\Delta_{o}=(\Delta_{S}+\Delta_{T})/\sqrt{2},$ while the order
parameter on the inner contour, $\Delta_{i}$, is $\Delta_{i}=(\Delta
_{S}-\Delta_{T})/\sqrt{2}.$ Within this definition, $\Delta_{S}$, is the order
parameter for a singlet pair and $\Delta_{T}$ is the order parameter for a
triplet pair. 
Note that the symbols $\Delta_i$, $\Delta_o$, $\Delta_S$ and $\Delta_T$ 
that are used in this discussion always refers to the superconducting order
parameter which can be different from the superconducting excitation gap, $\Delta$.
This picture implies four types of pairing interactions, which
corresponds to the following scattering processes of Cooper pairs:
$d_{o}\Longleftrightarrow d_{o},$ $d_{i}\Longleftrightarrow d_{i},$
$d_{o}\Longleftrightarrow d_{i},$ $d_{i}\Longleftrightarrow d_{o}.$ If
$\Delta_{o}=\Delta_{i},$ then in most (albeit not necessarily all) experiments
the triplet component cancels out. 
For example, such a situation can arise
following the considerations of Shaffer et al. \cite{shaffer2020crystalline}
where they take the intraband scattering within the same valley,
$d_{o,K}\Longleftrightarrow d_{o,K}$ or $d_{i,K}\Longleftrightarrow d_{i,K}$
to be the same (denoted as $g_{2}$ in Ref.~\cite{shaffer2020crystalline}),
which differs from their consideration of intraband scattering between the {\rm K} and
{\rm K}$^{\prime}$ valley, $d_{o,K}\Longleftrightarrow d_{i,K^{\prime}},$
$d_{i,K}\Longleftrightarrow d_{o,K^{\prime}}$ (denoted as $g_{3}$ in
Ref.~\cite{shaffer2020crystalline}).

If the pairing is due to phonons, then the matrix element for intraband
pairing interactions, $g_{oo}^{p}$ (or $g_{ii}^{p}$), between momenta
{\bf k} and {\bf p} is defined as,
\begin{align}
\left\langle d_{o,\mathbf{k}}|v|d_{o,\mathbf{p}}\right\rangle  &
=\left\langle \mathbf{k},o,\uparrow\right\vert \left\langle \mathbf{-k}%
,o,\downarrow\right\vert v\left\vert \mathbf{p},o,\uparrow\right\rangle
\left\vert -\mathbf{p},o,\downarrow\right\rangle \nonumber\\
&  =\left\langle \mathbf{k},o,\uparrow\right\vert g\left\vert \mathbf{p}%
,o,\uparrow\right\rangle \left\langle \mathbf{-k},o,\downarrow\right\vert
g\left\vert -\mathbf{p},o,\downarrow\right\rangle \nonumber\\
&  =g_{oo}^{p} \label{eq:goo,gii}%
\end{align}
where the phonon Green's function is included in $g$, The nondiagonal
electron-phonon matrix element that couples the outer and inner contours,
$g_{oi}^{p}$ is defined as,
\begin{align}
\left\langle d_{o,\mathbf{k}}|v|d_{i,\mathbf{p}}\right\rangle  &
=\left\langle \mathbf{k},o,\uparrow\right\vert \left\langle \mathbf{-k}%
,o,\downarrow\right\vert v\left\vert \mathbf{p},i,\downarrow\right\rangle
\left\vert -\mathbf{p},i,\uparrow\right\rangle \nonumber\\
&  =\left\langle \mathbf{k},o,\uparrow\right\vert g\left\vert -\mathbf{p}%
,i,\uparrow\right\rangle \left\langle \mathbf{-k},o,\downarrow\right\vert
g\left\vert \mathbf{p},o,\downarrow\right\rangle \nonumber\\
&  =g_{oi}^{p} \label{goi}%
\end{align}

Since scattering by phonons does not flip spin, electron-phonon coupling is
only relevant for $\left\vert K,\nu\right\rangle \Longleftrightarrow\left\vert
K,\nu^{\prime}\right\rangle $ scattering if $\nu=\nu^{\prime}$ or for
$\left\vert K,\nu\right\rangle \Longleftrightarrow\left\vert K^{\prime}%
,\nu^{\prime}\right\rangle $ if $\nu\neq\nu^{\prime}$. The indices $\nu$ and
$\nu^{\prime}$ can be either $o$ or $i$ to denote a state on an outer ($o$) or
inner ($i$) contour.

On the other hand, ferromagnetic spin-fluctuations, even though they can only
couple states withing the same valley, can work within both the inter- and
intraband channels. To determine the relative sign and strength of these
interactions, we define the amplitude of a spin fluctuation as $\mathbf{S}$,
and to a reasonable approximation their correlators can be assumed to be
spin-rotationally invariant: $\left\langle S_{\mathbf{k}}^{z}S_{\mathbf{p}%
}^{z}\right\rangle =\left\langle S_{\mathbf{k}}^{x}S_{\mathbf{p}}%
^{x}\right\rangle =\left\langle S_{\mathbf{k}}^{y}S_{\mathbf{p}}%
^{y}\right\rangle =\left\langle S_{\mathbf{k}}^{+}S_{\mathbf{p}}%
^{-}\right\rangle /2,$ leading to an isotropic spin susceptibility, $\chi$.
Within this definition, we can now describe the interaction of an electronic
state that crosses the Fermi level with a fluctuating spin moment as
$\mathbf{\sigma\cdot S}=\sigma^{z}S^{z}+(\sigma^{+}S^{-}+\sigma^{-}S^{+})/2.$
Hence, just as we did for the intraband electron-phonon interaction
(Eq.~\ref{eq:goo,gii}), we can write down the following expression for the
intraband interaction due to spin fluctuations, $g_{oo}^{s}$ (or $g_{ii}^{s}%
$):
\begin{align}
\left\langle d_{o,\mathbf{k}}|\left\langle \mathbf{S\otimes S}\right\rangle
|d_{o,\mathbf{p}}\right\rangle  & =\left\langle \mathbf{k},o,\uparrow
\right\vert \sigma^{z}S^{z}\left\vert \mathbf{p},o,\uparrow\right\rangle
\nonumber\\
& \times\left\langle \mathbf{-k},o,\downarrow\right\vert \sigma^{z}%
S^{z}\left\vert -\mathbf{p},o,\downarrow\right\rangle \nonumber\\
g_{oo}^{s}  & =-\frac{1}{4}\left\langle \mathbf{k},o|\mathbf{p},o\right\rangle
\chi(\mathbf{k-p)}\label{gsoo}%
\end{align}
Note that the sign of $g_{oo}^{s}$ is negative, indicating intraband spin-fluctuation
mediated interactions are repulsive, as would occur in a singlet channel. We
also define an expression for spin-fluctuation mediated interactions that
couple the outer and inner contours, $g_{oi}^{s}$, as
\begin{align}
\left\langle d_{o,\mathbf{k}}|\left\langle \mathbf{S\otimes S}\right\rangle
|d_{i,\mathbf{p}}\right\rangle  & =-[\left\langle \mathbf{k},o,\uparrow
\right\vert \sigma^{+}S^{-}\left\vert \mathbf{p},i,\downarrow\right\rangle
\nonumber\\
& \times\left\langle \mathbf{-k},o,\downarrow\right\vert \sigma^{-}%
S^{+}\left\vert -\mathbf{p},i,\uparrow\right\rangle ]/4\nonumber\\
g_{oi}^{s}  & =-\frac{1}{2}\left\langle \mathbf{k},o|\mathbf{p},i\right\rangle
\chi(\mathbf{k-p)}\label{gsii}%
\end{align}
where the minus sign in this matrix element appears because of our phase
convention for $\Delta_{i}$. Notice the prefactor $g_{oi}^{s}$, is a factor
of two larger than $g_{oo}^{s}$.

Hence, intraband and interband interactions around a fixed valley ({\rm K} or
{\rm K}$^{\prime}$) have the same sign as expected for singlet pairs. However, they
also have distinct prefactors, which is as if the standard singlet rotational
factor of 3 has distributed itself in a ratio of 2:1 between the interband and
intraband contributions within the SO-split bands. This combination of pairing
interactions due to phonons and spin fluctuations is summarized in Table
\ref{tab:pairs}. We have provided an alternative and more complete derivation
of these interactions in Appendix B.\begin{table}[ptb]
\textrm{%
\begin{tabular}
[c]{cccc}\hline\hline
Pair & $g^{p}$ &  & $g^{s}$\\\hline
$\left\vert K,o\right\rangle $,$\left\vert K,o\right\rangle $ & g$_{oo}^{p}%
$$>$0 &  & g$_{oo}^{s}$$<$0\\[0.5ex]%
$\left\vert K,i\right\rangle $,$\left\vert K,i\right\rangle $ & g$_{ii}^{p}%
$$>$0 &  & g$_{ii}^{s}$$<$0\\[0.5ex]%
$\left\vert K,o\right\rangle $,$\left\vert K^{\prime},o\right\rangle $ & $0$ &
& $\approx0$\\[0.5ex]%
$\left\vert K,i\right\rangle $,$\left\vert K^{\prime},i\right\rangle $ & $0$ &
& $\approx0$\\[0.5ex]%
$\left\vert K,o\right\rangle $,$\left\vert K,i\right\rangle $ & 0 &  &
g$_{oi}^{s}$$<$0\\[0.5ex]%
$\left\vert K,o\right\rangle $,$\left\vert K^{\prime},i\right\rangle $ &
g$_{oi}^{p}$$>$0 &  & $\approx$0\\\hline\hline
\end{tabular}
}\caption{ Sign of the pairing interaction that involves either phonons,
$g^{p}$, or ferromagnetic spin fluctuations, $g^{s}$, between two states on
the outer and/or inner contour at {\rm K} and {\rm K}$^{\prime}$. The states involved in
pairing are denoted as $\left\vert \mathbf{k},\nu\right\rangle $ where
$\mathbf{k}$ is a state either at {\rm K} or {\rm K}$^{\prime}$, $\nu$ can take on an
index $o$ or $i$ to denote a state on the outer ($o$) or inner ($i$). The
pseudospin ($\uparrow$ or $\downarrow$) associated with states on each contour
is depicted schematically in Figure ~\ref{fig:k_schem}. Attractive pairing
interactions are positive and repulsive interactions are negative.}%
\label{tab:pairs}%
\end{table}

The line of reasoning that leads to $\Delta_{o}=\Delta_{i}$, is based on the
fact that the bands in question are two-dimensional and nearly parabolic, so
their DOS is essentially the same (which DFT calculations confirm), while the
direction of the spins is anti-aligned. If the strength of the pairing
interaction is similar, this hypothesis of $\Delta_{o}=\Delta_{i}$ is
confirmed, and the net superconducting order parameter exhibits a singlet
character (see also Appendix B). 
Note that while the difference in the DOS of the outer and inner contours
is vanishingly small, the $k_{F}$ splitting is definitely not negligible in NbSe$_{2}$.
In monolayers of other TMDs, such as TaS$_2$ where Ising superconductivity has been
observed the $k_{F}$ splitting is larger than NbSe$_2$ \cite{sergio2018tuning}.
The implications of this large splitting of $k_{F}$ on the strength
of the pairing interaction on either contour is currently not known.
 Thus in the following we ask; does the strength of the pairing interaction have to
be similar on the outer and inner contours?

First, we discuss intraband interactions. They have a pairing component due to
phonons that is determined by the Eliashberg function, $\alpha{^{2}%
}F(\mathbf{q},\omega)$, which will have a characteristic momentum, \textbf{q},
and energy, $\omega$, dependence, and a pair-breaking component due to
ferromagnetic spin fluctuations, which is determined by the \textbf{q}%
-dependent spin susceptibility, $\chi(\mathbf{q},\omega)$. The $q$-dependent
spin susceptibility is sharply peaked at $q\sim0$ (Fig.~\ref{fig:Jq}). The
$q$-dependence of the electron-phonon coupling likely has a non-negligible
\textbf{q}-dependence as well.

If the intraband interaction due to phonons is the same for states on the
outer and inner contours, $g_{oo}^{p}$ = $g_{ii}^{p}$ (an approximation
adopted within Ref.~\cite{shaffer2020crystalline}), then the ratio of the
superconducting gap values, $|\Delta_{o}|/|\Delta_{i}|,$ is inversely
proportional to the square root of the ratio of the density of states
\cite{mazin2003electronic} of the outer and inner contour, $\sqrt{N_{o}/N_{i}%
}$, which is, as we know, essentially 1. The phase, however, depends on the
net sign of the interaction: if $|g_{oi}^{p}|>|g_{oi}^{s}|$, the phase is the
same, and the net order parameter is essentially singlet. However, if
$|g_{oi}^{p}|<|g_{oi}^{s}|,$ which is feasible, the order parameter is net-triplet.

Let us now discuss the potential for parity mixing due to a given structure of the
intraband coupling. We assume that the net intraband interaction due to
electron-phonon, $g^{p}(\mathbf{q)}$, and spin-fluctuation mediated
interactions, $g^{s}(\mathbf{q)}$, is peaked at small $q$. To be specific, we
take this net interaction to have a Lorentzian dependence on \textbf{q} (it
may as well have a sharp minimum at \textbf{q}=0, or have some other
comparable structure within momentum space):
\begin{equation}
g(\mathbf{q=k-k}^{\prime})\propto\xi^{2}/(q^{2}+\xi^{2}) \label{eq:gpq}%
\end{equation}

The net intraband coupling constants, $g_{oo}$ and $g_{ii}$, can be obtained
by averaging Eq.~\ref{eq:gpq} over all $k$ and $k^{\prime}$ on a circular
Fermi contour with a radius $k_{F}$. If we now consider how $g(\mathbf{q})$
changes if $k_{F}$ changes by $\delta k_{F}$, where $\delta k_{F}$ is the
SO-coupling induced splitting of the Fermi contours, we find:
\begin{equation}
\frac{g_{ii}-g_{oo}}{g_{ii}+g_{oo}}=\frac{\delta k_{F}/k_{F}}{1+(\xi
/2k_{F})^{2}},
\end{equation}
Unlike the DOS, which for a parabolic two-dimensional band does not depend on
$k_{F}$, this expression for the net pairing interaction due to intraband
interactions depends linearly on $\delta k_{F}$ and thus on the magnitude of
the SO-coupling. If $\xi\gg2k_{F},$ it vanishes, but if $\xi\sim2k_{F},$ it
leads to a non-negligible correction. Note that in NbSe$_{2}$ $\delta
k_{F}/k_{F}\sim1/3,$ and in TaS$_{2}$ $\delta k_{F}/k_{F}\sim1/2$! While
momentum-resolved calculations of the electron-phonon coupling in monolayer
NbSe$_{2}$ are underway \cite{margine}, the qualitative considerations we have
presented above challenges the current notion that the superconducting order
parameter in monolayer NbSe$_{2}$ is purely singlet, and demonstrates the
order parameter can indeed host a measurable admixture of triplet character.
Furthermore, our consideration of possible pairing interactions in 
Table ~\ref{tab:pairs} makes it evident that due to the strong spin-orbit
interaction of NbSe$_2$,
equal-spin triplet state is not possible at zero magnetic field.

Finally, while this is not the main subject of our paper, we briefly comment
on the ramifications and plausible experimental probes of the singlet-triplet
mixing of the order parameter of monolayer NbSe$_{2}$. Indeed, a number of
recent studies have alluded to the possibility of singlet-triplet mixing of
the order parameter in monolayer NbSe$_{2}$
\cite{mockli2020ising,cho2020distinct,hamill2020unexpected} by invoking
extrinsic mechanisms such as impurities and strain. The discussion we have
presented above suggests this parity mixing of the order parameter can have an
intrinsic origin depending on the interplay between momentum-dependent phonon
and spin-fluctuation induced couplings. Experiments that attempt to elucidate
this parity mixing need to access the parity-dependent coherence factors. One
possibility is quasiparticle interference, where the main challenge is to
separate the intraband ($o-o$ and $i-i)$ scattering from the interband
scattering processes. Magneto-optical spectroscopy using microwaves in the
deep infrared region of the spectrum is another potential experimental probe.
Finally, in the spirit of Ref.~\cite{mockli2020ising}, one expects that
impurities may affect superconductivity differently, depending on the parity.
All of these probes require quantitative theories, that are beyond the scope
of this present paper. Our primary goal was to demonstrate that mixed parity
Ising superconductivity is possible in the transition metal dichalcogenides,
and we hope this will encourage further theoretical and experimental research
into its manifestation.

\section{Conclusions}

We have developed a formalism that adapts the model theory for Ising
superconductivity into first-principles DFT calculations. We demonstrated that
bulk and monolayer NbSe$_{2}$ are close to a magnetic instability, and
spin-fluctuation induced interactions cannot be neglected when addressing
superconductivity in NbSe$_{2}$. Finally, we outlined two parametrically
admissible situations where superconductivity in monolayer NbSe$_{2}$ may be
partially triplet or even predominantly triplet without invoking an external
magnetic field or exchange bias, and point to the need to reexamine the
symmetry of the order parameter in monolayer NbSe$_{2}$. This perspective on
the role of magnetism in monolayer NbSe$_{2}$ will also be crucial to
understand and control the superconducting properties of monolayer NbSe$_{2}$
in the presence of an external magnetic field or with heterostructures between
monolayer NbSe$_{2}$ and magnetic materials.
\newline

{\it Note added $-$ }After our manuscript was first submitted we became aware of a related
preprint by another group \cite{divilov2020interplay} that appeared after our initial submission, which
also indicates that bulk and monolayer NbSe$_2$ are close to a magnetic instability.

\acknowledgements

The authors thank Maxim Khodas, Roxana Margine and Rafael Fernandes for useful
discussions. D.W was supported by a National Research Council (NRC) fellowship
at the US Naval Research Laboratory. I.I.M. was supported by ONR through grant N00014-20-1-2345.
\appendix

\section{Computational Methods}

Our calculations are based on density functional theory within the
projector-augmented wave method \cite{Blochl_PAW} as implemented in the VASP
code \cite{VASP_ref,VASP_ref2} using the generalized gradient approximation
defined by the Perdew-Burke-Ernzerhof (PBE) functional
\cite{perdew1996generalized}. We found it is essential that Nb $5s^{1}%
,4s^{2},4p^{6},4d^{4}$ electrons and Se $4s^{2},4p^{4}$ electrons are treated
as valence. All calculations use a plane-wave energy cutoff of 400~eV.
We use a ($18\times18\times1$) $\Gamma
$-centered $k$-point grid for the monolayer structure and a ($18\times
18\times9$) $k$-point grid for the bulk structure when performing structural
optimization and calculating the electronic structure.
The cell shape, volume and atomic
positions of the bulk structure were optimized with the Grimme-D3 van der
Waals correction \cite{grimme2010consistent} using a force convergence
criteria of 5 meV/\AA.

To determine the spin susceptibility, $\chi$ we used collinear and non-collinear
fixed-spin moment (FSM) calculations (sometimes referred to as the constrained local moments
approach).  In our collinear FSM calculations we constrain the magnitude of the magnetic moment
on the Nb atom.
To determine $\chi$ along a given crystal axes
we use non-collinear FSM calculations.
In these calculations we constrain the direction (either parallel or perpendicular to the
$c$-axis) and the magnitude of the magnetic moment (varying from 0 $\mu_B$ to
0.8 $\mu_B$).  With these constraints applied we then apply a Lagrange
multiplier to the minimization of the total energy.  Performing
these minimizations as a function of increasing magnetic moment and along a given direction
allows us to determine the change in energy with respect to the non-magnetic ground state
as a function of the total magnetization, $m$, for the bulk and monolayer structures.  We then
fit our results to an expansion of the total energy as a function
of $m$ (Eq.~[2]) to determine $\chi$.

The spin susceptibility, $\chi$, obtained from FSM calculations is sensitive
to the choice in $k$-point grid density, the energy convergence threshold, occupation
method and the
number of magnetization values used in the fit to expansion in the total energy
as a function of magnetic moment.
To yield converged values of $\chi$ we found it is essential to use a
($28\times28\times1$) $\Gamma
$-centered $k$-point grid for the monolayer structure and a ($28\times
28\times14$) $k$-point grid for the bulk structure, an energy convergence
threshold of 10$^{-8}$ eV, and up to 50 energy versus magnetization points
between 0 $\mu_B$ and 0.8 $\mu_B$ for all of the FSM calculations.
We also found that improved convergence was
achieved using the tetrahedron method to determine total energies for
both the bulk and monolayer structure (despite the 2D nature of the electronic structure).

For the calculation of the exchange constants within the disordered local
moment (DLM) approximation \cite{gyorffy1985first}. We used the
Korringa--Kohn--Rostokker method within the atomic sphere approximation
\cite{ruban1999calculated} and the Green's function-based magnetic-force theorem
\cite{liechtenstein1987local}. The implementation of this technique has been
described elsewhere \cite{ruban2004atomic}. This technique can be
considered to be a magnetic analogue of the disordered alloys theory based on
the coherent potential approximation. Calculations were performed for 13
nearest neighbor coordination spheres for five values of the fixed Nb
moments, 0.05 $\mu_{B}$, 0.10 $\mu_{B}$, 0.15 $\mu_{B}$, 0.25 $\mu_{B}$, and 0.35 $\mu_{B}$,
and extrapolated to $M$=0.
The nearest neighbor 
ferromagnetic exchange constant is at least 10 times larger than any 
other, but the remaining RKKY interaction is extremely long range and is 
responsible for weak satellites in Fig.~\ref{fig:Jq}.
Charge self-consistency was achieved using 147 irreducible
k-points in the Brillouin zone, and then an extended set of k-points (2565) to
compute the exchange constants.

\section{Singlet and Triplet Pairing Interactions}

To provide an additional derivation of the interactions in
Sec.~\ref{sec:results-mixing}, it is useful to consider the contribution of
density (that is, electron-phonon) and spin interactions to pairing on the {\rm K}
and {\rm K}$^{\prime}$ Fermi surfaces. Here, we write these interactions as
\begin{equation}
\frac{1}{2}\sum_{q}\rho(q)n_{q}n_{-q}+\frac{1}{2}\sum_{i,q}J_{i}%
(q)S_{i,q}S_{i,-q} \label{Appen1}%
\end{equation}
where $n_{q}=\sum_{k,s}c_{k+q/2,s}^{\dagger}c_{k-q/2,s}$ and $S_{i,q}%
=\sum_{q,s,s^{\prime}}c_{k+q/2,s}^{\dagger}\sigma_{i,s,s^{\prime}}c_{k-q/2,s}%
$, and $\sigma_{i}$ is a Pauli matrix. For clarity, we have allowed the
spin-interaction $J_{i}(q)$ to depend upon spin direction $i$, and will later
impose isotropy $J_{i}(q)=J(q)$. Eq.~\ref{Appen1} assumes interactions take
the same form as when inversion symmetry is present, implying we only consider
inversion symmetry breaking through single particle interactions. Noting that
for sufficiently large Ising spin-orbit coupling, pairing will only occur
between states of opposite spin, the contribution of the above interaction
towards superconductivity can be written as
\begin{equation}%
\begin{split}
&  \sum_{k,k^{\prime}}\Big [\rho(k-k^{\prime})-J_{x}(k+k^{\prime}%
)-J_{y}(k+k^{\prime})-J_{z}(k-k^{\prime})\Big ]\\
&  c_{k,\uparrow}^{\dagger}c_{-k,\downarrow}^{\dagger}c_{-k^{\prime
},\downarrow}c_{k^{\prime},\uparrow}%
\end{split}
\end{equation}
where we have used $\rho(k)=\rho(-k)$ and $J_{i}(k)=J_{i}(-k)$. We now examine
Cooper pairs formed from Fermions near the {\rm K} and {\rm K}$^{\prime}$ points. To
this end we define operators
\begin{align}
d_{o}^{\dagger}(k)  &  =c_{K+\delta k_{o},\uparrow}^{\dagger}c_{K^{\prime
}-\delta k_{o},\downarrow}^{\dagger}\\
d_{i}^{\dagger}(k)  &  =-c_{K+\delta k_{i},\downarrow}^{\dagger}c_{K^{\prime
}-\delta k_{i},\uparrow}^{\dagger}%
\end{align}
where $\delta k_{o}$ ($\delta k_{i}$) denote a wavevector on the outer (inner)
Fermi pocket at the {\rm K} point. Here we have introduced the same sign convention for $d_{i}^{\dagger}$ 
as in Eq.~\ref{eq:d0,di} in the main text.
For these operators, we find intraband,
$g_{ii}$ and $g_{oo}$, and interband, $g_{oi}$ and $g_{io}$, interactions due
to electron-phonon coupling and spin can be defined as:
\begin{align}
&  \sum_{k,k^{\prime}}\Big [g_{ii}(k,k^{\prime})d_{i}^{\dagger}(k)d_{i}%
(k^{\prime})+g_{oo}(k,k^{\prime})d_{o}^{\dagger}(k)d_{o}(k^{\prime})\Big ]\\
&  \sum_{\delta k,\delta k^{\prime}}\Big [g_{io}(k,k^{\prime})d_{i}^{\dagger
}(k)d_{o}(k^{\prime})+g_{oi}(k,k^{\prime})d_{o}^{\dagger}(k)d_{i}(k^{\prime
})\Big ]
\end{align}
where
\begin{align}
g_{ii}(k,k^{\prime})=  &  \rho(\delta k_{i}-\delta k_{i}^{\prime})-J(\delta
k_{i}-\delta k_{i}^{\prime})\nonumber\\
&  -2J(Q+\delta k_{i}+\delta k_{i}^{\prime})
\end{align}%
\begin{align}
g_{oo}(k,k^{\prime})=  &  \rho(\delta k_{o}-\delta k_{o}^{\prime})-J(\delta
k_{o}-\delta k_{o}^{\prime})\nonumber\\
&  -2J(Q+\delta k_{o}+\delta k_{o}^{\prime})
\end{align}%
\begin{align}
g_{io}(k,k^{\prime})  &  =g_{oi}(k,k^{\prime})=\rho(Q+\delta k_{i}+\delta
k_{o}^{\prime})\nonumber\\
&  -J(Q+\delta k_{i}+\delta k_{o}^{\prime})-2J(\delta k_{i}-\delta
k_{o}^{\prime}),
\end{align}
where $Q=2K$ and we have imposed spin-isotropy $J_{i}(k)=J(k)$. From this
expression, and taking $J(Q+\delta k)\approx0$, the coupling constants found
in Table \ref{tab:pairs} can be readily deduced.

It is instructive to consider the limit $\delta k_{i}=\delta k_{o}%
\rightarrow0$, then, when $J(Q)\approx0$, $g_{ii}=g_{oo}=-g^{p}(0)+g^{s}(0)$
and $g_{io}=g_{oi}=-g^{p}(Q)+2g^{s}(0)$ where the constants $g^p(0)$ and $g^p(Q)$are 
defined to be positive, corresponding to attractive electron-phonon interactions, and  $g^s(0)$ 
is negative corresponding to repulsive ferromagnetic interactions.  In this case, a pure singlet state
corresponds to the operator $[d_{i}(k)+d_{o}(k)]/\sqrt{2}$ for which the interaction is
\begin{equation}
v_{s}=-g^{p}(0)-g^{p}(Q)-3g^{s}(0).
\end{equation}
A pure triplet case corresponds to the operator $[d_{i}(k)-d_{o}(k)]/\sqrt{2}$, for which the
interaction is
\begin{equation}
v_{t}=-g^{p}(0)+g^{p}(Q)+g^{s}(0).
\end{equation}
These expressions reveal how spin-fluctuations strongly suppress the
spin-singlet state and enhance the spin-triplet state. 
Notice that once the spin fluctuations become sufficiently strong, 
that is $|g^p(Q)|<2|g^s(0)|$, the triplet solution will have a higher $T_c$ than the singlet solution. 

\section{Critical field anisotropy for a nodal Fermi surface}

\label{sec:app-gamma} As we discuss in Sec.~\ref{sec:results-mag} the third
Fermi pocket, around $\Gamma$, has zero SOC splitting along the $\Gamma-$M and
$\Gamma-$M$^{\prime}$ directions. Here, we rederive the expression for the
spin susceptibility for this band topology that accounts for the nodes along
these directions.

Assuming that the SO-splitting varies angularly as $\lambda\cos(3\varphi),$ we
derive the the susceptibility, $dm/dH_{x}$, for an in-plane magnetic field
applied along $\hat{x}$.
\begin{equation}
m =\frac{H}{2\pi\lambda}\int_{0}^{2\pi}\frac{d\varphi}{\sqrt{\cos^{2}%
\varphi+\zeta^{2}}}%
\end{equation}
where $\lambda$ is the maximal SOC splitting on this Fermi contour and
$\zeta=H/\lambda$. This gives the same Pauli expression as before, but with a
logarithmic correction:%
\begin{equation}
\frac{dm}{dH_{x}}=\chi_{\rm Pauli}\left(1+\frac{3}{4}\zeta^{2}\log\zeta\right)
\end{equation}
In the superconducting state, $\lambda\cos(3\varphi)$ is replaced by
$\sqrt{\lambda^{2}\cos^{2}(3\varphi)+\Delta^{2}},$ where $\Delta$ is the
average superconducting gap along the $\Gamma$ contour and $\zeta$ defined
above is replaced with $\zeta=\sqrt{\Delta^{2}+H^{2}}/\lambda$. Then in the
superconducting state
\begin{equation}%
\begin{split}
m  &  =\frac{1}{2\pi}\int_{0}^{2\pi}\frac{Hd\varphi}{\sqrt{\lambda^{2}\cos
^{2}\varphi+\Delta^{2}+H^{2}}}\\
&  =\frac{m}{2\pi\lambda}\int_{0}^{2\pi}\frac{d\varphi}{\sqrt{\cos^{2}%
\varphi+\zeta^{2}}}\\
\end{split}
\end{equation}
Upon integration, we\ get a logarithmic correction to the susceptibility in
the superconducting state, namely%
\begin{equation}
\frac{dm}{dH}=\chi_{\rm Pauli}\left(1+\frac{1}{4}\zeta_{0}^{2}\log\zeta_{0}\right)
\end{equation}
where $\zeta_{0} = \Delta/\lambda$. That is to say, the anisotropy of the
thermodynamic critical field is not infinite, but, roughly,
\begin{equation}
\left\vert \frac{N_{\Gamma}+N_{K}}{N_{\Gamma}\zeta_{0}^{2}\log\zeta_{0}%
}\right\vert ,
\end{equation}
where $N_{\Gamma(K)}$ is the DOS around the $\Gamma$ contour and $N_{K}$ is
the total DOS around the {\rm K} and {\rm K}$^{\prime}$ pockets. While this factor is
formally finite, it is a very large number of the order of 10$^{3}.$ 
Instead, other factors, such as a substrate induced Rashba spin-orbit coupling \cite{he2018magnetic}
and impurity scattering \cite{mockli2020ising}, are more important in limiting the anisotropy of the critical field.

\newpage\ \newpage

\end{document}